\documentstyle[amsmath,a4,11pt]{article}

\date{}

\begin{document}

\title{{\bf Cosmic magnetism, curvature and the expansion
dynamics}}

\author{David R. Matravers\thanks{e-mail address:
david.matravers@port.ac.uk} and Christos G. Tsagas\thanks{e-mail
address: christos.tsagas@port.ac.uk}\\ {\small Relativity and
Cosmology Group, Division of Mathematics and Statistics,}\\
{\small Portsmouth University, Portsmouth~PO1~2EG, England}}

\maketitle

\begin{abstract}
We discuss how a cosmological magnetic field could affect the
expansion of the universe, through its interaction with the
spacetime geometry. The tension of the field lines means that the
magneto-curvature coupling tends to accelerate positively curved
regions and decelerate those with negative curvature. Depending on
the equation of state of the matter, the effect varies from
mimicking that of a cosmological constant to resembling a
time-decaying quintessence. Most interestingly, the coupling
between magnetism and geometry implies that even weak fields have
a significant impact if the curvature contribution is strong. This
leads to kinematical complications that can inhibit the onset of
the accelerated phase in spatially open inflationary models. We
employ a simple cosmological model to illustrate these effects and
examine the conditions necessary for them to have a appreciable
impact.\\\\ PACS number(s): 98.80.Hw, 04.40.Nr, 95.30.Qd, 98.62.En
\end{abstract}

\section{Introduction}
Current observations provide strong evidence for the widespread
presence of magnetic fields in the universe. Magnetic fields
appear to be a common property of the intracluster medium of
galaxy clusters, extending well beyond the core regions \cite{K}.
Strengths of ordered magnetic fields in the intracluster medium of
cooling-flow clusters exceed those typically associated with the
interstellar medium of the Milky Way, suggesting that galaxy
formation and even cluster dynamics are, at least in some cases,
influenced by magnetic forces. Furthermore, reports of Faraday
rotation associated with high redshift Lyman-$\alpha$ absorption
systems seem to imply that dynamically significant magnetic fields
may be present in condensations at high redshift \cite{KPZ}. In
summary, the more we look for extragalactic magnetic fields, the
more ubiquitous we find them to be.

The origin of cosmic magnetism remains a mystery and is still a
matter of debate. Over the years, a number of possible solutions
has been proposed, ranging from eddies and density fluctuations in
the early plasma to cosmological phase-transitions, inflationary
and superstring inspired scenarios \cite{V}. Historically, studies
of magnetogenesis were motivated by the need to explain the origin
of large-scale galactic fields. Typical spiral galaxies have
magnetic fields of the order of a few $\mu$G coherent over the
plane of their disc. Such fields could arise from a relatively
large primordial seed field, adiabatically amplified by the
collapse of the protogalaxy, or by a much weaker one that has been
strengthened by the galactic dynamo. Provided that this mechanism
is efficient, the seed can be as low as $\sim10^{-23}$G at
present. However, in the absence of nonlinear dynamo
amplification, seeds of the order of $10^{-12}$G or even
$10^{-8}$G are required \cite{Ku}.

Magnetic fields introduce new ingredients into the standard, but
nevertheless uncertain, picture of the early universe. A
fundamental and unique property of magnetic fields is their
vectorial nature, which couples the field to the spacetime
geometry via the Ricci identity (see Eq. (\ref{3gcl})). An
additional, also unique, characteristic is the tension (i.e. the
negative pressure) exerted along the field's lines of force. This
means that every small magnetic flux tube behaves like an
infinitely elastic rubber band \cite{P}. Intuitively, what the
magneto-curvature coupling does, is to inject these elastic
properties of the field into space itself. The implications of
such an interaction are kinematical as well as dynamical with
quite unexpected results. Kinematically speaking, the
magneto-curvature effect tends to accelerate positively curved
perturbed regions, while it decelerates regions with negative
local curvature \cite{TB}. Dynamically, the most important
magneto-curvature effect is that it can reverse the pure magnetic
effect on density perturbations. To be precise, in the absence of
curvature, the field is found to slow down the growth of density
gradients \cite{TB}. However, when curvature is taken into
account, the inhibiting magnetic effect is reduced and, in the
case of `maximum' spatial curvature contribution, even reversed
\cite{TM}. Here, we focus upon the kinematics and provide an
example of how a cosmological magnetic field can modify, through
its coupling to geometry, the expansion rate of an almost-FRW
universe.

We assume a spacetime filled with a perfectly conducting
barotropic fluid and permeated by a weak primordial magnetic
field. The energy density and the anisotropic pressure of the
field are treated as first-order perturbations upon the FRW
background. The vectorial nature of the field results in a
magneto-geometrical term in the Raychaudhuri equation, which
depends on the curvature of the unperturbed model. The negative
pressure, that is the tension, carried by the magnetic force-lines
makes the implications of this term unique. Qualitatively speaking
the effect depends on the curvature sign of the background
spacelike sections. When the unperturbed universe is spatially
open, the magneto-geometrical term adds to the decelerating effect
of ordinary matter. For a spatially closed background, however,
the magneto-curvature contribution tends to accelerate the
expansion. In both cases the magnetic tension brings the expansion
rate closer to that of a flat FRW model. Quantitatively, the
effect depends on the relative strength of the field and on the
type of matter that fills the universe. In particular, the
magneto-geometrical term remains constant throughout an epoch of
stiff-matter domination. During this period the field acts as an
effective (positive or negative) cosmological constant. As the
universe progresses into the radiation  and subsequently the dust
era, the magneto-curvature term progressively decreases mimicking
a time-decaying quintessence. Under normal circumstances these
effects are relatively weak, although subtle enough to make an
open FRW universe look less open and a closed one look less
closed. On the other hand, the magneto-curvature effect on the
expansion can be dramatic, if the field or the curvature are
strong. Moreover, even weak magnetic fields have a significant
overall impact in a strongly curved universe. In fact, the mere
magnetic presence in spatially open inflationary models leads to
kinematical complications that can suppress the onset of the
accelerated phase. In a particular example, weakly magnetised FRW
universes with $p=-\rho$ cannot enter a period of de Sitter
inflation as long as $\Omega<0.5$.

In what follows, we employ the covariant perturbation formalism
\cite{EE, EB} (applied to magnetised cosmologies in \cite{TB,TM})
to illustrate the kinematical implications of cosmological
magnetic fields. The reader is referred to the aforementioned
articles for further discussion and details.

\section{Magneto-curvature effects on the expansion}
We consider a perturbed, slightly inhomogeneous and anisotropic,
FRW universe filled with a single perfectly conducting barotropic
medium with energy density $\rho$. We also allow for a magnetic
field ($B_a$), which is weak relative to the dominant matter
component (i.e. $B^2=B_aB^a\ll\rho$). The magnetic field is
assumed to be a test-field on the FRW background. The energy
density ($\rho_{\rm mag}={\textstyle{1\over2}}B^2$), the isotropic
pressure ($p_{\rm mag}={\textstyle{1\over6}}B^2$) and the
anisotropic stresses ($\pi_{ab}=-B_{\langle a}B_{b\rangle}$) of
the field will be treated as first-order
perturbations.\footnote{Angled brackets denote the projected,
symmetric, trace-free part of tensors and the orthogonal
projections of vectors. The same notation is also used for the
orthogonally projected time derivatives.} Note that the spatial
hypersurfaces of the unperturbed model may be closed or open as
well as flat. Thus, the zero-order Friedmann equation is given by
\begin{equation}
\frac{3k}{a^2}=8\pi G\rho_0- 3H^2+ \Lambda\,, \label{bFe}
\end{equation}
where $k=0,\,\pm1$ is the curvature index of the spatial
hypersurfaces, $\rho_0$ is the background matter density,
$H=\dot{a}/a$ is the Hubble parameter ($a$ is the scale factor)
and $\Lambda$ the cosmological constant.

In the background $H={\textstyle{1\over3}}\Theta$, where $\Theta$
describes the rate of the (volume) expansion and obeys
Raychaudhuri's formula. In its non-linear form, the latter reads
\cite{TB}
\begin{equation}
\dot{\Theta}+ {\textstyle{1\over3}}\Theta^2+ 4\pi
G\left(\rho+3p+B^2\right)- {\rm D}^aA_a- A_aA^a+ 2\left(\sigma^2-
\omega^2\right)- \Lambda=0\,, \label{Ray}
\end{equation}
where $p$ is the fluid pressure, $A_a$ is the 4-acceleration and
$\sigma^2$, $\omega^2$ are respectively the shear and vorticity
magnitudes. Note that ${\rm D}_a$ is the covariant derivative
operator projected orthogonally to the fluid flow. The crucial
magnetic effects propagate through the fluid acceleration, which
satisfies the non-linear Euler equation \cite{TB}
\begin{equation}
\left(\rho+p+{\textstyle{2\over3}}B^2\right)A_a+ c_{\rm s}^2{\rm
D}_a\rho+ \varepsilon_{abc}B^b{\rm curl}B^c+ A^b\pi_{ba}=0\,,
\label{Aa}
\end{equation}
where $c_{\rm s}^2$ is the sound speed of the barotropic fluid.
Linearising the divergence ${\rm D}^aA_a$ and using the
commutation law between the  projected gradients of spacelike
vectors, namely \cite{TB, EB}
\begin{equation}
{\rm D}_{[a}{\rm D}_{b]}B_c= {\textstyle{1\over2}}{\cal
R}_{dcba}B^d- \varepsilon_{abd}\omega^d\dot{B}_{\langle
c\rangle}\,, \label{3gcl}
\end{equation}
we can calculate the first-order magnetic contribution to Eq.
(\ref{Ray}). Note that formula (\ref{3gcl}), also known as the
3-Ricci identity, is the source of the magneto-curvature coupling
discussed here. It illustrates the vectorial nature of the field
and leads inevitably to curvature-dependent terms every time the
gradients of the magnetic vector commute. Here, it is written in
its exact, fully non-linear, form. Thus, $\omega_a$ and ${\cal
R}_{abcd}$ are respectively the vorticity vector and the `spatial'
Riemann tensor of the real spacetime.

Expressed in terms of the deceleration parameter and given the
weakness of the field, the linearised Raychaudhuri equation
becomes
\begin{eqnarray}
{\textstyle{1\over3}}\Theta^2{\rm q}&=&4\pi G\rho(1+3w)-
\frac{2kc_{\rm a}^2}{(1+w)a^2}+ \frac{c_{\rm
s}^2\Delta}{(1+w)a^2}+ \frac{c_{\rm a}^2{\cal
B}}{2(1+w)a^2}\nonumber\\&{}&- \frac{2}{\rho(1+w)}\left[\left({\rm
D}_{\langle a}B_{b\rangle}\right)^2- \left({\rm
D}_{[a}B_{b]}\right)^2\right]- \Lambda\,, \label{Ray1}
\end{eqnarray}
where ${\rm q}=-1-3\dot{\Theta}/\Theta^2$ is the deceleration
parameter, $w=p/\rho$ and $c_{\rm a}^2=B^2/\rho$ is the Alfv\'en
speed. In deriving Eq. (\ref{Ray1}) we have used the fact that
$D^aB_a=0$, and employed the zero-order expressions ${\cal
R}_{ab}={\textstyle{1\over3}}{\cal R}h_{ab}$ and ${\cal R}=6k/a^2$
for the spatial Ricci tensor and Ricci scalar respectively (recall
that ${\cal R}_{ab}={\cal R}^c{}_{acb}$). Also, the scalars
$\Delta=(a^2/\rho){\rm D}^2\rho$ and ${\cal B}=(a^2/B^2){\rm
D}^2B^2$ represent fluctuations in the matter and the magnetic
energy densities respectively.

Clearly, the sign of the right-hand side of Eq. (\ref{Ray1})
determines the state of the expansion. Negative terms accelerate
the universe, while positive ones slow the expansion down. Note
the quantities $({\rm D}_{\langle
a}B_{b\rangle})^2={\textstyle{1\over2}}{\rm D}_{\langle
a}B_{b\rangle}{\rm D}^{\langle a}B^{b\rangle}$ and $({\rm
D}_{[a}B_{b]})^2={\textstyle{1\over2}}{\rm D}_{[a}B_{b]}{\rm
D}^{[a}B^{b]}={\textstyle{1\over2}}({\rm curl}B_a)^2$. They
respectively describe what one might call `magnetic-shear' and
`magnetic-vorticity' effects \cite{PE}. Interestingly, the impact
these magnetically induced terms have on the expansion is opposite
to that of their kinematic counterparts. Indeed, unlike the
kinematic shear, the magnetic-shear term in Eq. (\ref{Ray1}) is
negative and therefore tends to accelerate the universe. Also, the
curl of the field vector causes further gravitational collapse, in
direct contrast to the effects of ordinary kinematic vorticity.
Both terms, however, are quadratic with respect to the
field-vector gradients, which suggests that they only become
important in highly inhomogeneous situations. On these grounds, we
may ignore the magnetically induced shear and vorticity and
rewrite Eq. (\ref{Ray1}) as
\begin{equation}
{\textstyle{1\over3}}\Theta^2{\rm q}=4\pi G\rho(1+3w)-
\frac{2kc_{\rm a}^2}{(1+w)a^2}+ \frac{c_{\rm
s}^2\Delta}{(1+w)a^2}+ \frac{c_{\rm a}^2{\cal B}}{2(1+w)a^2}-
\Lambda\,. \label{Re}
\end{equation}
Locally, the first-order scalars $\Delta$ and ${\cal B}$ are
either positive or negative, depending on whether the perturbed
region is respectively over-dense or under-dense. On average,
however, one expects that $\Delta=0={\cal B}$. On the other hand,
the mean $\rho$ and $c_{\rm a}^2$ are always positive. As a
result, the spatial average of Eq. (\ref{Re}) gives
\begin{equation}
{\textstyle{1\over3}}\Theta^2{\rm q}=4\pi G\rho(1+3w)-
\frac{2kc_{\rm a}^2}{(1+w)a^2}- \Lambda\,, \label{<Re>}
\end{equation}
with $\dot{\rho}=-(1+w)\Theta\rho$ and $(c_{\rm
a}^2)^{\cdot}=-{\textstyle{1\over3}}(1-3w)\Theta c_{\rm a}^2$ ,
given that $(B^2)^{\cdot}=-{\textstyle{4\over3}}\Theta B^2$ to
first order. Thus, the coupling between the magnetic field and the
background spatial curvature affects the average deceleration of a
perturbed magnetised FRW universe. Qualitatively, the effect
depends on the geometry of the spacelike hypersurfaces. In
particular, when $k=-1$, the magneto-curvature term in Eq.
(\ref{<Re>}) simply adds to the gravitational pull of the
(ordinary) matter component. On the other hand, for a spatially
closed background the magneto-curvature coupling tends to
accelerate the expansion, thus opposing the matter effect. In both
cases the overall result of the field presence is to bring the
expansion rate closer to that of a flat universe. This
unconventional behaviour is caused by the negative pressure
experienced along the magnetic lines of force, that is by the
field's tension. The latter tends to smooth out the kinematic
effects of curvature, imprinted in Eq. (\ref{bFe}), by modifying
the expansion rate of the universe accordingly. Intuitively, one
might argue that the magneto-curvature coupling has transferred
the elastic properties of the field into space itself \cite{TB,
TM}. Note that for $k=+1$ the field will reverse the fluid effect
on the expansion, if the magneto-geometrical term in Eq.
(\ref{<Re>}) is stronger than the matter term. This is possible
when the field is strong or when the spatial regions are strongly
curved. Even a moderate magneto-curvature contribution, however,
can boost the expansion rate of a closed FRW universe and make it
look less closed, or slow down an open one to make it look less
open.

How important the above effects are and what period in the
lifetime of the universe they affect most, depends on the type of
the matter that fills the universe. As we shall see next, for
cosmological models with ordinary matter (i.e. $0\leq w\leq1$),
the most intriguing magneto-curvature effects occur in a $k=+1$
model. On the other hand, if exotic matter dominates (e.g. $-1\leq
w\leq-{\textstyle{1\over3}}$), the magneto-curvature coupling is
crucial when $k=-1$. This case also offers an example of how a
relatively weak magnetic field can have a strong impact.

\section{The magnetic field as an effective cosmological constant}
Equation (\ref{<Re>}) raises the interesting question as to
whether the magneto-curvature term can mimic a cosmological
constant. The answer is positive, depending on the matter
component of the universe. To be precise, the magneto-geometrical
term in Eq. (\ref{<Re>}) is time-independent if $c_{\rm
a}^2\propto a^2$. Given that $c_{\rm a}^2\propto a^{3w-1}$ this
happens when $w=1$. Thus, provided that stiff matter dominates,
the magnetic field introduces an effective $\Lambda$-term through
its coupling to the background curvature. Such an effective
cosmological constant is positive when the unperturbed FRW
universe is positively curved and negative if $k=-1$. Here, we
will focus upon the $k=+1$ case because then the magnetic effects
on the expansion oppose those of the matter. We set $\Lambda=0$
and consider an early period with $p=\rho$. Then Eq. (\ref{<Re>})
becomes
\begin{equation}
{\textstyle{1\over3}}\Theta^2{\rm q}= 16\pi G\rho- \frac{c_{\rm
a}^2}{a^2}\,,  \label{s<Re>}
\end{equation}
where $c_{\rm a}^2\propto a^2$ since $\rho\propto a^{-6}$. Note
how the magneto-curvature term acts as a positive cosmological
constant, having effectively replaced $\Lambda$. This term leads
to exponential expansion as long as it dominates the right-hand
side of Eq. (\ref{s<Re>}). On using expression (\ref{bFe}), Eq.
(\ref{s<Re>}) gives
\begin{equation}
{\textstyle{1\over3}}\Theta^2{\rm q}=
6H^2\Omega\left[1-\frac{\left(\Omega_0-1\right)c_{\rm
a}^2}{6\Omega}\right]\,,  \label{s<Re>1}
\end{equation}
where $\Omega\equiv\rho/\rho_{\rm c}$,
$\Omega_0\equiv\rho_0/\rho_{\rm c}$ are respectively the average
and background density parameters, with $\rho_{\rm
c}\equiv3H^2/8\pi G$ representing the background critical density.
For a weak magnetic field $\rho\simeq\rho_0$ on average, which
means that $\Omega\simeq\Omega_0$. On these grounds, we will no
longer distinguish between $\Omega$ and $\Omega_0$ but use them
interchangeably. Hence, the expansion is accelerated if
\begin{equation}
c_{\rm a}^2>\frac{6\Omega}{\Omega-1}\,.  \label{sac}
\end{equation}
This implies that marginally closed universes, with
$0<\Omega-1\ll1$, require very strong magnetic fields to
accelerate. When $\Omega-1\sim1$, however, a cosmological field
with energy density comparable to that of the stiff matter could
trigger a period of accelerated expansion. Stiff-matter FRW models
are encountered in the so called pre-Big-Bang scenarios as the
dual counterparts of the string-theory inspired dilaton
cosmologies. They also correspond to scalar-field models dominated
by the field's kinetic energy \cite{Ve}.

One should keep in mind that as the magnetic field gets stronger
the almost-FRW treatment given here becomes less reliable. In this
respect, condition (\ref{sac}) should only be taken as indicative.
Having said that, studies of perturbed magnetised Bianchi~I models
have shown that, qualitatively speaking, the magnetic effects on
average scalars (such as $\Theta$) remain very close to those
predicted by the FRW treatments (see Eq. (68) in \cite{TM}).

In Eq. (\ref{s<Re>}) the magneto-geometrical term drops slower
than the matter term, which means that it can accelerate the
expansion later in the stiff-matter era. Using the evolution law
$\rho=\rho_*(1+z)^6$, we find that the acceleration starts at
redshift
\begin{equation}
z\simeq-1+\sqrt[6]{\frac{(c_{\rm
a}^2)_*\left(\Omega_*-1\right)}{6\Omega_*}}\,, \label{sz}
\end{equation}
where for convenience $z_*=0$ at the end of the stiff-matter era
rather than today.

\section{The magnetic field as quintessence}
Let us now consider a radiation dominated universe. When
$w={\textstyle{1\over3}}$ Eq. (\ref{<Re>}) becomes
\begin{equation}
{\textstyle{1\over3}}\Theta^2{\rm q}=8\pi G\rho- \frac{3c_{\rm
a}^2}{2a^2}\,, \label{r<Re>}
\end{equation}
with $\rho\propto a^{-4}$ and $c_{\rm a}^2={\rm constant}$. The
magneto-curvature term is no longer constant but decreases as
$a^{-2}$, slower than the matter term. Here, the
magneto-geometrical effects resemble those attributed to a
time-decaying quintessence \cite{CDS}. As before, Eqs. (\ref{bFe})
and (\ref{r<Re>}) imply that the accelerated phase commences if
$c_{\rm a}^2>2\Omega/(\Omega-1)$ at redshift
$z\simeq-1+\sqrt{(c_{\rm a}^2)_0(\Omega_0-1)/2\Omega_0}$, the
latter measured at the time of matter-radiation equality. Although
magnetic fields of such strength are not allowed at
nucleosynthesis, \cite{COST}, they are not a priori excluded
earlier in the radiation era. Neutrino damping means that
relatively strong magnetic fields in the early radiation era can
efficiently dissipate their energy to satisfy the nucleosynthesis
limits \cite{JKO} .

Qualitatively speaking, the picture does not change in the dust
era. The difference now is that the magneto-curvature term drops
as fast as the fluid term (i.e. $\propto a^{-3}$). Similarly to
the radiation era, the field mimics a time-decaying quintessence
and leads to accelerated expansion if $c_{\rm
a}^2>3\Omega/4\left(\Omega-1\right)$. Such magnetic fields,
however, are beyond the limits set by current observations
\cite{BFS}. Nevertheless, even weak fields can slightly accelerate
the expansion and thus make a magnetised closed FRW universe look
less closed than it actually is. Note that when $k=-1$ only the
sign of the effects discussed so far changes. In spatially open
models the magneto-curvature effects on the expansion are
complementary to those of the ordinary matter.

\section{Strong effects from weak magnetic fields}
So far we have restricted ourselves to magnetised cosmologies
filled with ordinary matter (i.e. $0\leq w\leq1$). In these
environments the magneto-curvature effects, subtle though they may
be, remain secondary unless the field is relatively strong.
However, strong magnetic fields are not always necessary for the
magneto-curvature effect to be significant. In fact, the
aforementioned interaction between magnetism and geometry may also
challenge the widespread perception that magnetic fields are
relatively unimportant for cosmology. This belief is based on
current observations, which point towards a weak magnetic presence
at nucleosynthesis and recombination. However, the
magneto-curvature coupling could make the field into a key player
irrespective of the magnetic strength. In principle, even weak
magnetic fields can lead to appreciable effects, provided that
there is a strong curvature contribution. To illustrate how this
can happen, we turn to spatially open cosmological models
containing matter with negative pressure. For $k=-1$ Eq.
(\ref{<Re>}) becomes
\begin{equation}
{\textstyle{1\over3}}\Theta^2{\rm q}=4\pi G\rho(1+3w)+
\frac{2c_{\rm a}^2}{(1+w)a^2}\,, \label{<-1Re>}
\end{equation}
where we have set $\Lambda=0$. We begin with a simple qualitative
argument. In Eq. (\ref{<-1Re>}) the magneto-curvature term evolves
as $c_{\rm a}^2/a^2\propto a^{-3(1-w)}$ and the matter term obeys
the standard evolution law $\rho\propto a^{-3(1+w)}$. Thus,
$(c_{\rm a}^2/a^2)/\rho\propto a^{6w}$, which implies that the
magneto-geometrical effects dominate the right hand side of Eq.
(\ref{<-1Re>}) at sufficiently early times if $w<0$. Note that the
Alfv\'en speed, which measures the relative strength of the field,
behaves as $c_{\rm a}^2\propto a^{-1+3w}$. This means that, as we
go back in time, the ratio $(c_{\rm a}^2/a^2)/\rho$ grows faster
than the Alfv\'en speed provided that $w<-{\textstyle{1\over3}}$.
Therefore, when $-1< w<-{\textstyle{1\over3}}$, the
magneto-geometrical effects can dominate the dynamics of the early
expansion while the field is still relatively weak. In these cases
the phase of accelerated expansion, which otherwise would have
been inevitable, may not happen. Instead, the universe goes
through a period of decelerated expansion. This is an interesting
possibility that puts a question mark on the efficiency of
inflationary models in the presence of primordial magnetism.
Recall that an initial curvature era was never considered as a
problem for inflation, given the smoothing power of the
accelerated phase. However, this may not be the case when a
primordial magnetic field is present, no matter how weak the
latter is.

There is a plethora of scenarios, which utilise out-of-equilibrium
epochs in the early universe to generate primeval magnetic fields
\cite{V}. The energy scales involved vary from $\sim100$ MeV at
the QCD phase transition, to $\sim100$ GeV in the case of
electro-weak (EW) physics and closer to the Planck energy scale
for inflation or string cosmology. The viability of the proposed
mechanisms depends primarily on the field's subsequent evolution,
in view of the current observational constraints. Crucially for
our purposes, only a weak (seed) magnetic field is required. As we
shall see next, when curvature dominates, it is the mere presence
of the field that is important and not its relative strength. Note
that  in the current section we assume a magnetic presence at
epochs earlier than the EW phase transition, when the
SU(2)$\times$U(1)$_{Y}$ symmetry is restored. If primordial
magnetic fields were to be present at these high temperatures,
they should correspond to the U(1)$_{Y}$ hypercharge rather than
U(1)$_{EM}$. Such hyper-electromagnetic fields have the same
stress-tensor of their EM counterparts and, in an infinitely
conducting medium, the remaining hyper-magnetic field obeys an
induction equation and satisfies a vanishing-divergence law
analogous to the standard Maxwell equations (see e.g. \cite{GS}).
Thus, the formulae derived in Sec. 2 are also compatible with
epochs prior to the EW symmetry-breaking.

Equation (\ref{<-1Re>}) can also apply to weakly magnetised,
almost-FRW, scalar-field dominated cosmologies. Indeed, given the
absence of a background magnetic field (see Sec. 2), consider a
FRW unperturbed model containing a self-interacting complex scalar
field $\phi$. Relative to a timelike 4-velocity $u_a$, the stress
tensor associated with $\phi$ has the perfect-fluid form
$T_{ab}=\rho u_au_b+ph_{ab}$, where
$\rho={\textstyle{1\over2}}\dot{\phi}\dot{\phi}^*+V(\phi\phi^*)$
and $p={\textstyle{1\over2}}\dot{\phi}\dot{\phi}^*-V(\phi\phi^*)$
\cite{SJ}. On the other hand, the magnetic field always behaves as
an imperfect fluid with $T_{ab}={\textstyle{1\over2}}B^2u_au_b+
{\textstyle{1\over6}}B^2h_{ab}-B_{\langle a}B_{b\rangle}$. Thus,
the total energy momentum tensor reads
\begin{equation}
T_{ab}=\left(\rho+{\textstyle{1\over2}}B^2\right)u_au_b+
\left(p+{\textstyle{1\over6}}B^2\right)h_{ab}- B_{\langle
a}B_{b\rangle}\,,  \label{tTab}
\end{equation}
assuming that, to leading order, the coupling between $\phi$ and
magnetism does not affect the perfect-fluid behaviour of the
scalar field.\footnote{Following scalar electrodynamics, a typical
Lagrangian coupling $\phi$ to the electromagnetic vector potential
$A_a$ has the form
\begin{equation}
{\cal L}=-{\textstyle{1\over2}}{\cal D}_a\phi({\cal D}^a\phi)^*-
V(\phi\phi^*)- {\textstyle{1\over4}}F_{ab}F^{ab}\,. \label{L}
\end{equation}
In the above $F_{ab}=2\nabla_{[a}A_{b]}$ is the Faraday tensor,
${\cal D}_a=\nabla_a-ieA_a$ is the gauge covariant derivative ($e$
is the electromagnetic coupling) and $V=V(\phi\phi^*)$ is the
potential that describes the self-interaction of $\phi$.} This is
a reasonable approximation given the weakness of the magnetic
field. Moreover, the pure geometrical nature of the
magneto-curvature interaction means that imperfections in the
fluid description of $\phi$ are of minor importance for our
purposes. The effects we are examining here are triggered solely
by the spacetime geometry and by the tension of the magnetic
force-lines. In addition to, their impact on the expansion
dynamics depends almost entirely on the strength on the background
curvature. On these grounds, one may substitute expression
(\ref{tTab}) into the conservation law $\nabla^bT_{ab}=0$, obtain
the formulae of Sec. 2 and eventually recover Eq. (\ref{<-1Re>}),
this time for a spatially open scalar-field dominated universe.

Let us now consider the implications of the magnetic presence for
a simple inflationary model. The initial conditions at the onset
of inflation are rather unclear and subject to debate. Usually,
the universe enters the inflationary regime from the Planck era or
after a highly relativistic epoch. The accelerated expansion is
driven by the dominating inflaton field $\phi$, with an equation
of state that satisfies the condition $-\rho\leq
p<-{\textstyle{1\over3}}\rho$. Here, we also allow for a weak
primeval magnetic field. To begin with, recall our qualitative
argument that as long as the index $w=p/\rho$ varies within
$(-1,\,-{\textstyle{1\over3}})$, there is always an early period
when the expansion is dominated by the magneto-curvature effects.
The latter could suppress the accelerated phase. To refine and
quantify this statement we employ Eq. (\ref{bFe}) and then (see
Sec. 3) rewrite Eq. (\ref{<-1Re>}) as
\begin{equation}
{\textstyle{1\over3}}\Theta^2{\rm
q}={\textstyle{3\over2}}H^2\left[(1+3w)\Omega+\frac{4c_{\rm
a}^2(1-\Omega)}{3(1+w)}\right]\,, \label{<-1Re>1}
\end{equation}
which reduces to the standard non-magnetised expression when the
field term is dropped \cite{ME}. Thus, the condition for
suppressing acceleration (i.e. for ${\rm q}>0$) reads
\begin{equation}
c_{\rm a}^2>-\frac{3(1+w)(1+3w)\Omega}{4(1-\Omega)}\,. \label{dec}
\end{equation}
In spatially open inflationary cosmologies, the density parameter
diverges from the boundary point ($a=0,\,\Omega=0$) towards the
$\Omega=1$ limit \cite{ME}. Such models go through a curvature
dominated early stage characterised by $\Omega\ll 1$. During this
period, a relatively weak magnetic field (with $c_{\rm a}^2\ll1$)
is capable of slowing the expansion down, as condition (\ref{dec})
shows. For example, when $w=-{\textstyle{2\over3}}$ condition
(\ref{dec}) reduces to\footnote{The value
$w=-{\textstyle{2\over3}}$ also corresponds to the effective
equation of state associated with network of infinite planar
domain walls. Similarly, $w=-{\textstyle{1\over3}}$ also
represents a network of infinitely extended cosmic strings
\cite{VS}.}
\begin{equation}
c_{\rm a}^2>\frac{\Omega}{4(1-\Omega)}\,, \label{dwdec}
\end{equation}
which is satisfied by a weak magnetic field provided $\Omega\ll1$.
This strong curvature requirement is considerably relaxed as
$w\rightarrow-1,\,-{\textstyle{1\over3}}$, since then the
right-hand side of (\ref{dec}) becomes arbitrarily small. Note
that the limit $w=-1$ corresponds to standard slow-roll inflation
with ${\textstyle{1\over2}}\dot{\phi}\dot{\phi}^*\ll V$. On the
other hand, at $w=-{\textstyle{1\over3}}$ we have, what one might
call, a `minimal inflation' scenario. In both cases condition
(\ref{dec}) is reduced to the requirement that $c_{\rm a}^2>0$.
This, in turn, suggests that the mere magnetic presence will
suppress the accelerated phase in any open universe with
$w=-1,\,-{\textstyle{1\over3}}$, irrespective of how strong its
spatial curvature is. This is not surprising given that at $w=-1$
the magneto-curvature impact on the expansion maximises, while at
$w=-{\textstyle{1\over3}}$ the gravitational effects take their
minimum value (see Eq. (\ref{<-1Re>})). One can refine these
results by recalling the, in the magnetic presence, the total
gravitational mass of the universe is $\rho+3p+B^2$ instead of
$\rho+3p$ and the total energy density is
$\rho+p+{\textstyle{2\over3}B^2}$ rather than $\rho+p$. Given the
weakness of the field, this correction is of very little
importance when $-1<w<-{\textstyle{1\over3}}$. When
$w=-{\textstyle{1\over3}}$, however, the total gravitational mass
is $B^2$ and still positive, which explains why the model
decelerates. Of particular interest is the behaviour near the
$w=-1$ limit. In this case one should replace $1+w$ in condition
(\ref{dec}) with ${\textstyle{2\over3}}c_{\rm a}^2$ rather than
zero. Then, one obtains a refined, curvature depended, requirement
$\Omega<0.5$ for the suppression of the accelerated phase.
Therefore, moderately open universes dominated by a slowly rolling
scalar field cannot enter a period of de Sitter inflation as there
is an arbitrarily weak magnetic field present.

In the absence of cosmic magnetism, our model goes through an
accelerated phase until a more conventional equation of state is
restored, as $\phi$ rolls towards the minimum of its potential. In
the process, the curvature of the space is smoothed out and the
universe ends up arbitrarily flat. This apparent solution of the
flatness problem has long been considered a major point in favour
of the inflationary paradigm. However, by introducing even a weak
primordial magnetic field, one could drastically change this
picture. The results presented so far indicate that, in the
presence of primeval magnetism, inflation may not be able to cope
with negative curvature effectively.

In principle, one could still argue that our model might
eventually enter a late accelerated phase. Clearly, whether this
could happen or not depends on how strong the initial curvature is
and how long the inflaton-dominated regime lasts. The details rest
with the particular model that one might have in mind. However, it
seems plausible that the stronger the initial magneto-curvature
effects are, the longer the inflaton domination must be if the
universe is ever to accelerate. Thus, provided that the effective
index $w$ is trapped within $(-1,\,-{\textstyle{1\over3}})$ long
enough, the aforementioned magneto-curvature effects may
eventually become too weak to suppress inflation. On the other
hand, if the universe remains inflaton-dominated for a brief spell
only, the accelerated phase will never have the chance to begin.

Let us emphasise that the magneto-curvature effects discussed
above should be seen as the field's kinematic reaction to the
geometry of the spatial sections rather, than as a direct attempt
to suppress inflation. In an open universe the tension of the
magnetic force-lines slows down the expansion rate to bring it
closer to that of a flat FRW model (see discussion in Sec. 2). The
stronger the curvature is the more dramatic the effect. In this
respect, the magnetic presence does not generically target
inflation although it might seem so at first. In fact, as one can
immediately see through Eqs. (\ref{<Re>}) and (\ref{<-1Re>}), the
field would have assisted the inflationary expansion if the space
had been closed (i.e. if $k=+1$) instead of open.

\section{Discussion}
Despite their established widespread presence, research on
cosmological magnetic fields remains rather marginal. The reasons
could be the perceived weakness of the field effects or the lack,
as yet, of a consistent theory explaining the origin of cosmic
magnetism. The fact that magnetic fields further complicate the
picture of the early universe may be an additional factor.
However, magnetic fields have been observed everywhere where
modern technology has made their detection possible. Thus, we feel
justified to argue that a magnetic-free picture of the early
universe is not a complete picture. Moreover, we believe that the
potential of some unique magnetic characteristics, such as their
vectorial nature and tension properties, has not been fully
appreciated. The magneto-geometrical interaction discussed here is
a consequence of these properties and of the general relativistic
geometrical interpretation of gravity. Intuitively, what the
magneto-curvature coupling does, is to inject the elastic
properties of the field into space itself. The resulting effects
are unexpected and potentially very important. The main task of
this paper was to draw attention to these issues.

Our examples illustrate the impact of this coupling between
magnetism and geometry on the evolution of the universe. We have
discussed how, depending on the circumstances, the
magneto-curvature effects mimic those of a positive cosmological
constant, or those attributed to a time-decaying quintessence. For
a spatially closed background, the overall magnetic impact varies
from weakly opposing deceleration to accelerated expansion
depending on the field's strength. The stronger the field is, the
more dramatic the effect. On the other hand, when the background
is open, the magnetic contribution to the expansion is simply
complementary to that of ordinary matter. Even then, however, the
field's role is subtle, making the universe look less open than it
actually is.

The most intriguing result, however, is that even weak magnetic
fields can become, through their coupling to geometry, key players
in the evolution of the universe. We argue that, when the
curvature is strong, the mere presence of a magnetic field leads
to effects that can alter the picture of the universe in
unexpected ways. In particular, we have demonstrated how spatially
open cosmological models containing matter with negative pressure
are not guaranteed a period of early accelerated expansion if a
magnetic field is present. In fact, for $p=-\rho$ the mere
presence of the field can suppress the inflationary phase even in
moderately curved spaces with $\Omega<0.5$. Strong curvature is
important when $-\rho< p<-{\textstyle{1\over3}}\rho$, if
appreciable deceleration is to be achieved. Still, this effect can
be triggered by arbitrarily weak fields. The magnetic strength is
not the issue any more. Once the vectorial nature of the field has
brought geometry into play, the overall magnetic impact no longer
depends on the field alone. It is the presence of the magnetic
field that is important and not its relative strength.

\section*{Acknowledgements}
CGT was supported by PPARC. The authors would like to thank R.
Maartens, M. Bruni, D. Wands, A. Kandus, C. Ungarelli and B.
Bassett for helpful discussions and also the referee for
constructive comments.


\begin{thebibliography}{99}

\bibitem[1]{K} P.P. Kronberg, Rep. Prog. Phys.
{\bf 57}, 325 (1994)
\bibitem[2]{KPZ} P.P. Kronberg, J.J. Perry and E.L.H. Zukowski,
Ap. J. {\bf 387}, 528 (1992); A.M. Wolfe, K.M. Lanzetta and A.L.
Oren, Ap. J. {\bf 388}, 17 (1992)
\bibitem[3]{V} E.R. Harrison, M.N.R.A.S. {\bf 147}, 279
(1970); C. J. Hogan, Phys. Rev. Lett. {\bf 51}, 1488 (1983); M.S.
Turner and L.M. Windrow, Phys. Rev. D {\bf 30}, 2743 (1988); T.
Vachaspati, Phys. Lett. B {\bf 265}, 258 (1991); M. Gasperini, M.
Giovanini and G. Veneziano, Phys. Rev. Lett. {\bf 75}, 3796
(1995); G. Sigl, K. Jedamzik and A.V. Olinto Phys. Rev. D {\bf
55}, 4582 (1997); M. Joyce and M. Shaposhnikov, Phys. Rev. Lett.
{\bf 79}, 1193 (1997); A. Kandus, E.A. Calzetta, F.D. Mazzitelli
and C.E.M. Wagner, Phys. Lett. B {\bf 472}, 287 (2000)
\bibitem[4]{Ku} R.M. Kulsrud, in {\it Galactic and Extragalactic
Magnetic Fields}, edited by R. Beck, P.P. Kronberg, and R.
Wielebinski (Reidel, Dordrecht, 1990).
\bibitem[5]{P} E.N. Parker, {\it Cosmical Magnetic Fields}
(Oxford: Oxford University Press, 1979)
\bibitem[6]{TB} C.G. Tsagas and J.D. Barrow, Class. Quantum
Grav. {\bf 14}, 2539 (1997); C.G. Tsagas and J.D. Barrow, Class.
Quantum Grav. {\bf 15}, 3523 (1998); C.G. Tsagas and R. Maartens,
Phys. Rev. D {\bf 61}, 083519 (2000)
\bibitem[7]{TM} C.G. Tsagas and R. Maartens, Class. Quantum Grav. {\bf
17}, 2215 (2000)
\bibitem[8]{EE} J. Ehlers, Abh. Mainz Akad. Wiss. Lit.
{\bf 11}, 1 (1961); G.F.R. Ellis, in {\it Carg\`ese Lectures in
Physics} vol VI, edited by E. Schatzmann (New York: Gordon and
Breach 1973) p. 1; R. Maartens, Phys. Rev. D {\bf 55}, 463 (1997);
G.F.R. Ellis and H. van Elst, in {\it Theoretical and
Observational Cosmology}, edited by M. Lachi\'eze-Rey (Dordrecht:
Kluwer 1999) p. 1
\bibitem[9]{EB} S.W. Hawking, Ap. J. {\bf 145}, 544 (1966); G.F.R.
Ellis and M. Bruni, Phys. Rev. D {\bf 40}, 1804 (1989); G.F.R
Ellis, M. Bruni and J. Hwang, Phys. Rev. D {\bf 42}, 1035 (1990)
\bibitem[10]{PE} D. Papadopoulos and F.P. Esposito, Ap. J. {\bf 257},
10 (1982)
\bibitem[11]{Ve} G. Veneziano, Phys. Lett. B {\bf 265}, 287
(1991); M. Gasperini and G. Veneziano, Astropart. Phys. {\bf 1},
317 (1993); P.J.E. Peebles and A. Vilenkin, Phys. Rev. D {\bf 59},
063505 (1999)
\bibitem[12]{CDS} R.R. Caldwell, R. Dave and P.J. Steinhardt, Phys.
Rev. Lett. {\bf 80}, 1582 (1998)
\bibitem[13]{COST} B. Cheng, A.V. Olinto, D. Schramm and J. Truran,
Phys. Rev. D {\bf 54}, 4714 (1996); P. Kernan, G. Starkman and T.
Vachaspati, Phys. Rev. D {\bf 54}, 7202 (1996)
\bibitem[14]{JKO} K. Jedamzik, V. Katalinic and A.V. Olinto, Phys.
Rev. D {\bf 57}, 3264 (1998)
\bibitem[15]{BFS} J.D. Barrow, P.G. Ferreira and J. Silk, Phys.
Rev. Lett. {\bf 78}, 3610 (1997); P. Blasi, S. Burles and A.V.
Olinto, Ap. J. {\bf 514}, L79 (1999)
\bibitem[16]{GS} M. Giovannini and M. Shaposhnikov, Phys.
Rev. D {\bf 57}, 2186 (1998); R. Brustein and D.H. Oaknin, Phys.
Rev. Lett. {\bf 82}, 2628 (1999); R. Brustein and D.H. Oaknin,
Phys. Rev. D {\bf 60}, 023508 (1999)
\bibitem[17]{SJ} D. Scialom and P. Jetzer, Phys. Rev. D {\bf 51},
5698 (1995)
\bibitem[18]{ME} M.S. Madsen and G.F.R Ellis, M.N.R.A.S. {\bf
234}, 67 (1988)
\bibitem[19]{VS}A. Vilenkin and P. Shellard, {\em Cosmic Strings
and Other Topological Defects} (Cambridge: Cambridge University
Press 1994)

\end{thebibliography}
\end{document}